\documentclass[aps,reprint,superscriptaddress]{revtex4-1}
\usepackage{amssymb,amsmath,bm}
\usepackage[dvipdfmx]{graphicx}

\usepackage{color}
\usepackage{ulem}

\begin{document}


\title{Manipulating magnetoelectric effect -- Essence learned from Co$_4$Nb$_2$O$_9$}


\author{Yuki Yanagi}
\affiliation{Department of Physics, Meiji University, Kawasaki 214-8571, Japan}
\author{Satoru Hayami}
\affiliation{Department of Physics, Hokkaido University, Sapporo 060-0810, Japan}
\author{Hiroaki Kusunose}
\affiliation{Department of Physics, Meiji University, Kawasaki 214-8571, Japan}

\date{\today}

\begin{abstract}
 Recent experiments for  
 linear magnetoelectric (ME) response in honeycomb antiferromagnet Co$_4$Nb$_2$O$_9$ revealed that 
 the electric polarization can be manipulated by the in-plane rotating magnetic field in a systematic way.
  We propose the minimal model by extracting essential ingredients of Co$_4$Nb$_2$O$_9$ to exhibit
  such ME response.  
  It is the  
  three-orbital model with $xy$-type atomic spin-orbit coupling (SOC) on the single-layer honeycomb structure, 
   and  it is shown to reproduce qualitatively the observed field-angle dependence of the electric polarization.  
   The obtained results can be understood  by the perturbative calculation with respect to the atomic SOC.
   These findings could be useful to explore further ME materials having similar manipulability of the electric polarization.
\end{abstract}

\pacs{asdfasdf}

\maketitle


 The electrons in solids containing ions with partially-filled $d$- or $f$-shells have orbital degrees of freedom in addition to spin and charge ones. 
   Strong Coulomb repulsion between electrons with such multiple internal degrees of freedom generates many fascinating physics~\cite{Tokura}, 
    some of which have potential for novel electronic device applications, e.g., spintronics~\cite{Murakami,Hasan} and valleytronics~\cite{Rycerz}.       
 The magnetoelectric (ME) effect is a classical example of spin-charge-orbital coupled physics~\cite{Curie,Dzyaloshinskii,Astrov} and nonlinear ME effects have attracted 
  much attention owing to the discovery of the multiferroic compounds showing huge ME response~\cite{Kimura,Katsura,Mostovoy,Sergienko,Cheong,Khomskii,Arima}.

The linear  ME effect  has  also gained renewed interest 
   in the context of the emergent odd-parity magnetic multipolar orderings~\cite{Fiebig,Spaldin,Iniguez,Malashevich,Scaramucci,Picozzi,Yanase,Watanabe,Hayami,Hayami2,Hayami3,Kato}. 
 In   linear ME materials, a proper structure of the ME tensor $\hat{\alpha}$  
determines the magnetic(electric)-field controllability of the linear electric (magnetic) polarization. 
  For instance, in an archetypal ME compound, Cr$_2$O$_3$,  
   the ME tensor is    diagonal, i.e.,   $\alpha_{xx}=\alpha_{yy}\neq\alpha_{zz}$~\cite{Dzyaloshinskii}.  
  In this case, the ME response is longitudinal.  
     On the other hand, in Ni$_3$B$_7$O$_{13}$~\cite{Ascher}, the magnetic point group implies that 
     the only $\alpha_{yz}$ and $\alpha_{zy}$ components can be finite, which yields the transverse ME response in the $yz$-plane.

         Recently, Khanh \textit{et al.} have found the peculiar ME response  
         in honeycomb antiferromagnet Co$_4$Nb$_2$O$_9$, where the electric polarization is rotated by the in-plane rotating magnetic field \textit{with twice faster and in opposite direction}~\cite{Khanh,Khanh3}.
         However, the microscopic minimal conditions for such ME response remain unclear.
         Motivated by these observations, we elucidate minimal conditions to emerge such ME response by extracting essential ingredients of Co$_{4}$Nb$_{2}$O$_{9}$.
This could be useful to explore efficiently further ME materials having similar manipulability of the electric polarization.
In this paper, we first demonstrate that the minimal three-orbital model indeed exhibits the observed behavior of the electric polarization.
Then, we discuss the essential ingredients which can be related to some aspects of the original model for Co$_4$Nb$_2$O$_9$.
Lastly, we show that the obtained results can be understood by the perturbative calculation with respect to the atomic SOC.

\begin{figure*}[t]
\begin{center}
\includegraphics[width=16cm]{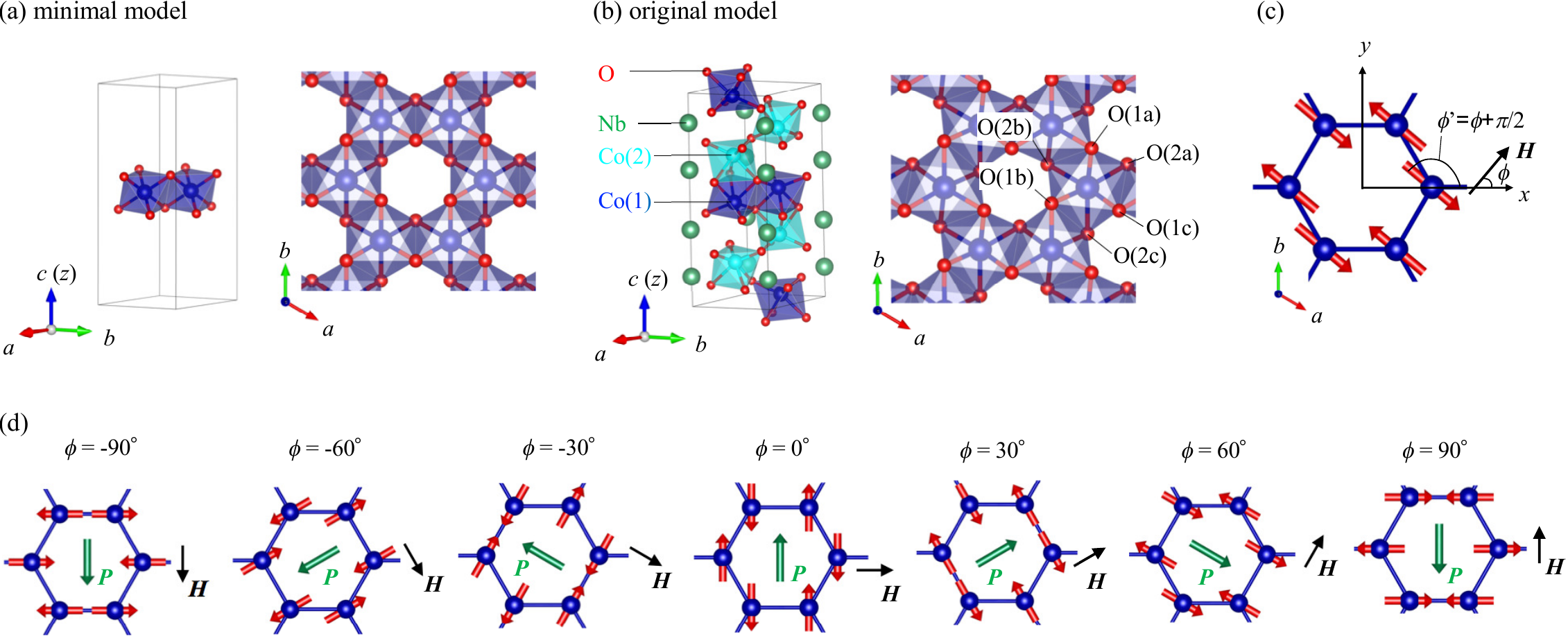}
\vspace*{6mm}
\caption{
The lattice structures of (a) our minimal model and (b) original model for Co$_4$Nb$_2$O$_9$. (c) The AFM structure under the in-plane magnetic field $\bm{H}$. 
 Note that $\bm{M}_\mathrm{AFM}$ is parallel to the $x$-axis in the absence of $\bm{H}$. 
(d) Schematic illustration of the induced  electric polarization $\bm{P}$ and 
the AFM moment $\bm{M}_{\mathrm{AFM}}$ by rotating magnetic field $\bm{H}$ in the $xy$-plane.  
 }
 \label{fig_model}
\end{center}
\end{figure*}

It has long been known that Co$_4$Nb$_2$O$_9$ shows linear ME effects in the antiferromagnetic (AFM) state~\cite{Fischer},
and the lattice structure is shown in Fig.~\ref{fig_model}(b)~\cite{Bertaut,Castellanos}.  
According to the recent neutron diffraction measurements for single crystals~\cite{Khanh,Khanh2} and powder samples~\cite{Denga}, 
the magnetic moments on Co-atoms are almost lying in the $xy$-plane and aligned antiferromagnetically in each honeycomb layer. 
These AFM honeycomb layers are stacked ferromagnetically along the $c$-axis. 
This AFM ordering breaks both spatial inversion and time-reversal symmetries, and it makes linear ME effects  
possible below the N\'eel temperature, $T_{\rm N}=27.2$K. 

Recent experimental reinvestigation 
revealed the  ME response of Co$_4$Nb$_2$O$_9$ in detail~\cite{Khanh,Khanh2,Khanh3,Fang,Solovyev}.  
 Due to  weak in-plane magnetic anisotropy,   
the AFM moment $\bm{M}_{\mathrm{AFM}}$ is almost always perpendicular to the in-plane external field $\bm{H}=H(\cos\phi,\sin\phi)=H(\sin\phi',-\cos\phi')$,  
 where $\phi'=\phi+\pi/2$ 
is the angle of $\bm{M}_{\mathrm{AFM}}$ measured from  the $x$-axis [see Fig.~\ref{fig_model}(c)].  
 Figure~\ref{fig_model}(d) depicts the induced electric polarization in the rotating magnetic field, which is characterized by $\bm{P}\sim P[\sin(-2\phi'),-\cos(-2\phi')]$. 
 From these observations, we can deduce the corresponding ME tensor in the form, 
\begin{multline}
\hat{\alpha}(\phi')
\propto r 
\begin{pmatrix}
 -\cos\phi'  & \sin\phi' \\
 \sin\phi'  & \cos\phi'
\end{pmatrix}
\\
+(1-r)
\begin{pmatrix}
 \cos 3\phi'  & \sin 3\phi' \\
 -\sin 3\phi'  &  \cos 3\phi'
\end{pmatrix},
\label{eq_me_exp}
\end{multline}
 where  
$r$ is an arbitrary  constant  independent of the angle $\phi'$, and the $z$ components are omitted. 
In our previous study~\cite{Yanagi}, we have successfully reproduced  
the observed ME response on the basis of the realistic 
model derived from the density functional band calculation. 
However, the essential ingredients  for such ME response remain unclear.

Let us begin with the minimal  three-orbital model with $xy$-type SOC 
on the two-dimensional honeycomb lattice under the AFM molecular field.
 This model corresponds to the simplified one, in which we take into account the partially filled three orbitals $\ell=e_{g1},e_{g2},a_{1g}$ and 
 the only single honeycomb layer composed of Co(1)O$_6$ octahedra in the original model for Co$_4$Nb$_2$O$_9$, and neglect the buckling structure [see Figs.~\ref{fig_model}(a) and (b)].
The Hamiltonian is given by
\begin{align}
\mathcal{H}=\mathcal{H}_{\mathrm{kin}}+\mathcal{H}_{\mathrm{SOC}}+\mathcal{H}_{\mathrm{AFM}}.
\label{eq_h}
\end{align}   
Each term of $\mathcal{H}$ is explicitly given as follows, 
\begin{align}
&
\mathcal{H}_{\mathrm{kin}} = \sum_{\bm{k}\sigma}\sum_{\alpha\beta\ell\ell'} H^{(0)}_{\alpha\ell\beta\ell'}(\bm{k}) d^{\dagger}_{\bm{k}\alpha\ell\sigma}d^{}_{\bm{k}\beta\ell'\sigma}, \\
&
\mathcal{H}_{\mathrm{SOC}} = \frac{\lambda}{2}\sum_{\bm{k}\alpha}\sum_{\ell\ell'\sigma\sigma'}(l_{\ell\ell'}^{x}\sigma^{x}_{\sigma\sigma'}+l_{\ell\ell'}^{y}\sigma^{y}_{\sigma\sigma'}) d^{\dagger}_{\bm{k}\alpha\ell\sigma}d^{}_{\bm{k}\beta\ell'\sigma'},
\label{eq_hsoc} \\
&
\mathcal{H}_{\mathrm{AFM}} = -\Delta\sum_{\bm{k}\alpha\ell}\sum_{\sigma\sigma'}p[\alpha]M_{\sigma\sigma'}(\phi') 
d^{\dagger}_{\bm{k}\alpha\ell\sigma}d^{}_{\bm{k}\alpha\ell\sigma'}, \nonumber \\
&\quad\quad\quad
\hat{M}(\phi')=\cos\phi' \hat{\sigma}^{x}+\sin\phi' \hat{\sigma}^{y}, \label{eq_hafm}
\end{align}
where $d^{(\dagger)}_{\bm{k}\alpha\ell\sigma}$ represents the annihilation (creation) operator for the electron on the sublattice $\alpha(=\mathrm{A},\mathrm{B})$ 
 with wave vector $\bm{k}$, orbital $\ell$ and spin $\sigma(=\uparrow,\downarrow)$, and $\hat{\sigma}^{i}$ ($i=x,y$) represents $i$-th component of the Pauli matrix.
$H^{(0)}_{\alpha\ell\beta\ell'}(\bm{k})$ is the kinetic energy including the crystalline-electric-field (CEF) potential and nearest-neighbor hopping on the two-dimensional honeycomb lattice.
 Here, the Slater-Koster parametrization is used as $dd\sigma=-0.15\mathrm{eV}$, $dd\pi=(2/3)dd\sigma$, and $dd\delta=dd\sigma/6$.
The CEF splitting is set to be $\varepsilon_{e_{g}}-\varepsilon_{a_{1g}}=0.62\mathrm{eV}$.
The magnitudes of the SOC and the AFM molecular field are set as $\lambda=0.1\mathrm{eV}$ and $\Delta=2.0\mathrm{eV}$, respectively.
These values are estimated from the density functional band calculation for Co$_4$Nb$_2$O$_9$~\cite{Yanagi}.
The factor $p[\alpha]=+1~(-1)$ for $\alpha=\mathrm{A}~(\mathrm{B})$ in Eq.~(\ref{eq_hafm}) is used to represent the staggered order.
There are 3 electrons per Co$^{2+}$ ion, since we assumed that 4 of 7 electrons in Co$^{2+}$ ion occupy the lowest $\varepsilon_{g}'$ orbitals as will be discussed later.
By the sufficiently large AFM molecular-field term $\mathcal{H}_{\rm AFM}$, the system becomes insulating. 
The explicit forms of the orbitals and orbital angular-momentum operators are given by (\ref{eq_basis}) and (\ref{eq_lop}), respectively.
 By diagonalizing the Hamiltonian in Eq.~(\ref{eq_h}) at each $\bm{k}$, we obtain the energy bands $\varepsilon_{\bm{k}\zeta}$ and corresponding eigenvectors $|\bm{k}\zeta\rangle$ ($\zeta=1-12$).

We investigate the linear ME responses  of the model Hamiltonian in Eq.~(\ref{eq_h})  by means of the standard Kubo formula.
 Since the external magnetic field acts on both the spin and orbital magnetic moments, the ME tensor $\hat{\alpha}$  
 is a sum of the spin part $\hat{\alpha}^{S}$ and orbital part $\hat{\alpha}^{L}$, where $\hat{\alpha}^{L}$ 
  is expressed by  the correlation function between the velocity and orbital  magnetic moment, 
  $\hat{Q}^{L}(z)=\langle\langle V^{i}; L^{j}\rangle \rangle_z$,  as follows:
\begin{align}
\hat{\alpha}^{L}=\lim_{\omega\rightarrow 0}\frac{\hat{Q}^{L}(\omega+i\eta)-\hat{Q}^{L}(i\eta)}{i\omega}.
\label{eq_me_1}
\end{align}
 Similarly, $\hat{\alpha}^{S}$ is obtained by replacing $L^{j}$ with $2S^{j}$.
The velocity, orbital magnetic moment~\cite{Thonhauser} and spin magnetic moment operators are given by
\begin{align}
&V^{i}=\sum_{\bm{k}\sigma}\sum_{\alpha\beta\ell\ell'} v^{i}_{\alpha\ell\beta\ell'}(\bm{k}) d^{\dagger}_{\bm{k}\alpha\ell\sigma}d^{}_{\bm{k}\beta\ell'\sigma},
\\
&L^{i}=\sum_{\bm{k}\alpha\sigma}\sum_{\ell\ell'}l^{i}_{\ell\ell'} d^{\dagger}_{\bm{k}\alpha\ell\sigma}d^{}_{\bm{k}\alpha\ell'\sigma},  \\ 
&S^{i}=\frac{1}{2}\sum_{\bm{k}\alpha\ell}\sum_{\sigma\sigma'}\sigma^{i}_{\sigma\sigma'} d^{\dagger}_{\bm{k}\alpha\ell\sigma}d^{}_{\bm{k}\alpha\ell\sigma'},
\label{eq_s}
\end{align}
with $\hat{v}^{i}(\bm{k})=\partial\hat{H}^{(0)}(\bm{k})/\partial k_{i}$.
The ME tensor $\alpha^{L}_{ij}$ in Eq.~(\ref{eq_me_1}) is explicitly calculated as follows
\begin{align}
\alpha^{L}_{ij}=\frac{1}{iV}\sum_{\bm{k}\zeta\zeta'}
\frac{\langle \bm{k} \zeta |V^{i}| \bm{k} \zeta' \rangle 
\langle \bm{k} \zeta' | L^{j}| \bm{k} \zeta \rangle} 
{\varepsilon_{\bm{k}\zeta}-\varepsilon_{\bm{k}\zeta'}} 
\frac{f(\varepsilon_{\bm{k}\zeta})-f(\varepsilon_{\bm{k}\zeta'})}{\varepsilon_{\bm{k}\zeta}-\varepsilon_{\bm{k}\zeta'}+i\eta},
\label{eq_me_2} 
\end{align}
where 
 $f(\varepsilon)=(e^{(\varepsilon-\mu)/T}+1)^{-1}$ is the Fermi distribution function with chemical potential $\mu$.
By the above ME tensor, the induced electric polarization is expressed as
$P_{i}/H=\alpha_{ix}\sin\phi'-\alpha_{iy}\cos\phi'$.

\begin{figure}[t]
\begin{center}
\includegraphics[width=8cm]{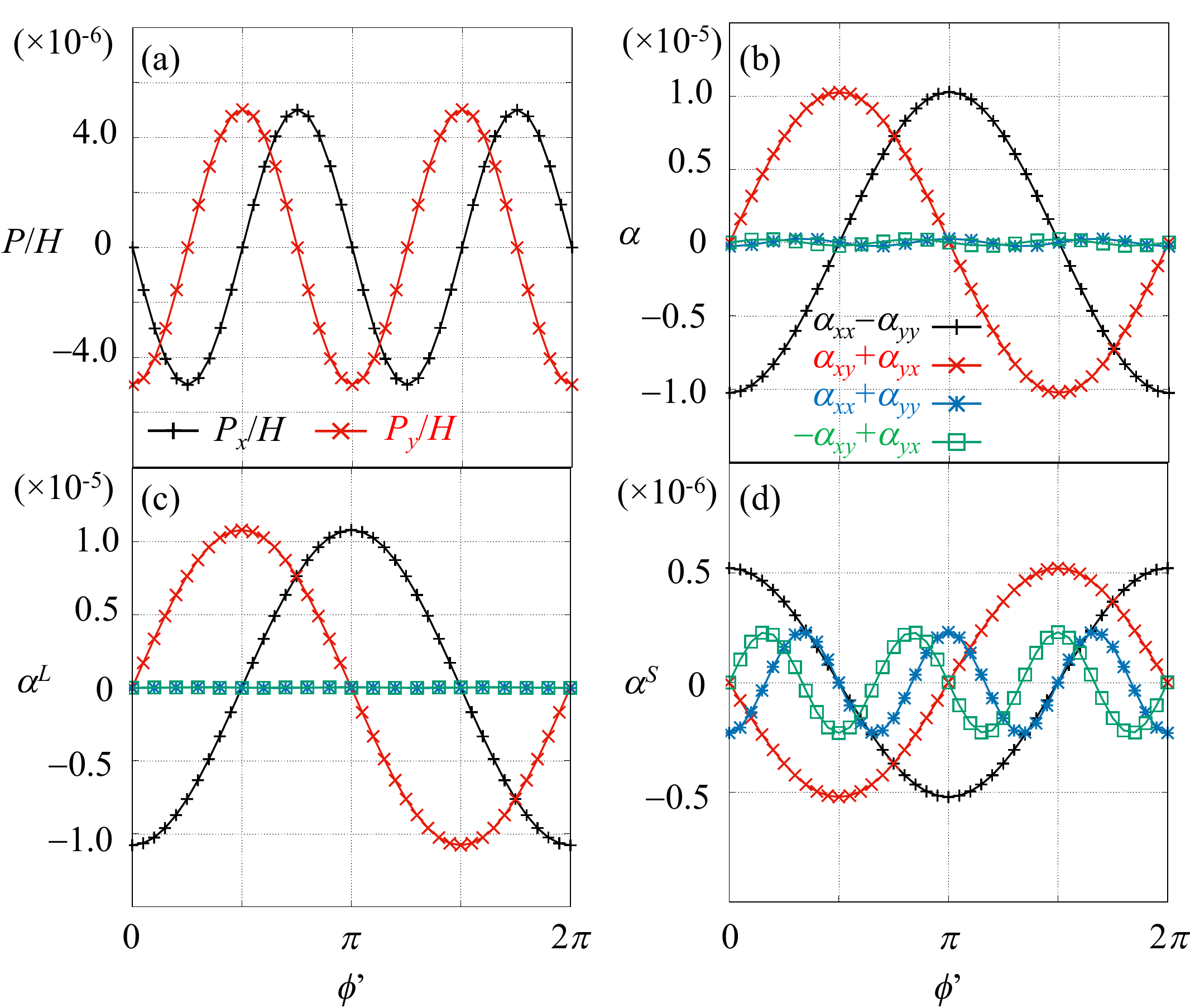}
\caption{Angle $\phi'$-dependence of $\bm{P}$ and $\hat{\alpha}$ at $T=0.01$eV. (a) electric polarization $\bm{P}$, 
(b) total ME tensor $\alpha_{ij}=\alpha^{S}_{ij}+\alpha^{L}_{ij}$, 
(c) orbital part $\alpha^{L}_{ij}$,  
and (d) spin part $\alpha^{S}_{ij}$.  
}
 \label{fig_ME}
\end{center}
\end{figure}

Figure~\ref{fig_ME}(a) shows the $\phi'$-dependence of the electric polarization $\bm{P}$. 
It is found that $\bm{P}\propto (-\sin 2\phi',-\cos 2\phi')$, which is consistent with the observed behavior. 
   The $\phi'$-dependences of $\alpha_{ij}$, $\alpha^{L}_{ij}$ and $\alpha^{S}_{ij}$
  are also shown in Figs.~\ref{fig_ME}(b), (c) and (d), respectively.  
In our minimal  model, $\hat{\alpha}^{L}$  dominates over $\hat{\alpha}^{S}$. 
 The orbital part $\hat{\alpha}^{L}$ has only the fundamental $\phi'$ rotation, which is characterized by 
  $r_{L}\sim 1$ in Eq.~(\ref{eq_me_exp}), while the spin part $\hat{\alpha}^{S}$ has both $\phi'$ and $3\phi'$ rotations characterized by $r_{S}\sim 0.0244$.
In total, $\hat{\alpha}$ is characterized by $r\sim 1.027$, indicating that magnetic quadrupoles play a dominant role in ME for Co$_4$Nb$_2$O$_9$.

 Next, we discuss the connection between our minimal model and the realistic model for Co$_4$Nb$_2$O$_9$. 
    In Co$_4$Nb$_2$O$_9$, O$^{2-}$-ions form a trigonally distorted octahedron  around a Co-atom as shown in Fig.~\ref{fig_model}(b).  
    The CEF from O$^{2-}$-ions splits  
    $3d$ orbitals of a Co-atom into the non-degenerate $a_{1g}$ orbital and two sets of doubly degenerate  
   $e_g$ and $e_g'$ orbitals.
 The wave functions of the $e_g$ and $a_{1g}$ orbitals are given by
\begin{align}
&|e_{g1}\rangle=  \frac{2}{\sqrt{6}}|yz\rangle - \frac{1}{\sqrt{3}}|x^2-y^2\rangle, \\
&|e_{g2}\rangle=   -\frac{2}{\sqrt{6}}|zx\rangle + \frac{1}{\sqrt{3}}|xy\rangle, \\
&|a_{1g}\rangle= |3z^2-r^2\rangle.
\label{eq_basis}
\end{align}
According to the first principles band calculations, 
    the CEF level scheme is as follows, $\varepsilon_{e_g}>\varepsilon_{a_{1g}}>\varepsilon_{e_g'}$, where $\varepsilon_{\ell}$ is the atomic energy of the orbital $\ell$~\cite{CEF_band}. 
  By considering the  electron configuration of Co$^{2+}$ ions, $(3d)^{7}$, we assume that the lowest lying $e_{g}'$ orbitals are fully occupied, 
  and the rest of $e_{g}$ and $a_{1g}$ orbitals are partially filled by 3 electrons.
Within these orbital space, the matrix elements of the orbital angular-momentum operators are given by 
\begin{align}
\hat{l}^{x}=\sqrt{2}\begin{pmatrix} 0 & 0 & -i \\ 0 & 0 & 0 \\ i & 0 & 0 \end{pmatrix},
\quad
\hat{l}^{y}=\sqrt{2}\begin{pmatrix} 0 & 0 & 0 \\ 0 & 0 & -i \\ 0 & i & 0 \end{pmatrix},
\label{eq_lop}
\end{align}
 and $\hat{l}^{z}$ vanishes.  
 As a result,  the SOC in our minimal model, Eq.~(\ref{eq_hsoc}), is the $xy$-type. 
 
Up to this point, the minimal ingredients to exhibit the in-plane ME response as Eq.~(\ref{eq_me_exp}) are given.
So that further differences between our minimal model and the realistic model for Co$_4$Nb$_2$O$_9$ do not play any important roles in the occurrence of the in-plane ME response.
We summarize the differences as follows.
In the original lattice structure, the unit cell contains four sets of the two distinct Co-atoms, Co(1) and Co(2) 
   and the edge-shared (corner-shared) Co(1)O$_6$  [Co(2)O$_6$] octahedra form  buckled honeycomb structures as shown in Fig.~\ref{fig_model}(b).  
   Here, we note  that in a  Co(1)O$_6$ octahedron, the triangle formed 
   by three O-atoms located on the upper plane of a Co(1)-atom,  O(1a)-O(1c),  
 and that formed by O-atoms on the lower plane, O(2a)-O(2c),  are not equivalent to each other as shown in Fig.~\ref{fig_model}(b).
In our minimal model,
 we assume that the upper triangle, O(1a)-O(1c), and the lower triangle, O(2a)-O(2c), are equivalent, and
  there are additional two-fold rotational symmetries along the nearest neighbor Co(1)-Co(1) bonds, 
[see Fig.~\ref{fig_model}(a)].
 Accordingly,  the point-group symmetry of the single Co(1) honeycomb layer in our minimal model is upgraded from
   the original $C_{3i}$ group to  
   higher  $D_{3d}$ group.

Finally, we discuss that the obtained ME response in our minimal model is  
naturally understood by the perturbative calculation with respect to the atomic SOC.
Likewise magnetic susceptibility, 
  one can calculate the correlation function $Q^{L(S)}_{ij}(i\nu_m)$
  by the Green's function technique in Matsubara framework, and  
   obtain $Q^{L(S)}_{ij}(\omega+i\eta)$  in Eq.~(\ref{eq_me_1}) by  
the analytic continuation procedure, $i\nu_m\rightarrow \omega+i\eta$, where $\nu_m=2m\pi T$ is the bosonic Matsubara frequency.
By the formal expansion of the non-perturbative Green's function $\hat{G}(k)$ in terms of $\hat{G}_{0}^{-1}(k)=(i\omega_{n}+\mu)\hat{1}-\hat{H}^{0}(\bm{k})$ 
with respect to the AFM molecular-field term, $-\Delta\hat{M}(\phi')\hat{\rho}^{z}$ with $\hat{\rho}^{z}$ being the $z$-component of the Pauli matrix in the sublattice space, we obtain
\begin{align}
\hat{G}(k)=&\hat{G}_{0}-\Delta \hat{G}_{0}\hat{M}\hat{\rho}^{z}\hat{G}_{0}+\Delta^{2} \hat{G}_{0}(\hat{M}\hat{\rho}^{z}\hat{G}_{0})^{2}-\cdots
\cr=&
\hat{G}_{0}[\hat{1}-\Delta^{2}(\hat{\rho}^z\hat{G}_{0})^2]^{-1} 
\cr&\quad
-\Delta \hat{M}\hat{G}_{0}\hat{\rho}^{z}\hat{G}_{0}[\hat{1}-\Delta^{2}(\hat{\rho}^z\hat{G}_{0})^{2}]^{-1}
\cr=&
\hat{G}_{\mathrm{E}}(k)+\hat{M}(\phi')\hat{G}_{\mathrm{O}}(k),
\label{eq_g}
\end{align}
where we have introduced the diagonal and off-diagonal Green's functions, $\hat{G}_{\rm E}$ and $\hat{G}_{\rm O}$, in the spin space as 
\begin{align}
&
\hat{G}_{\mathrm{E}}(k)=[\hat{G}_{0}^{-1}(k)-\Delta^{2}\hat{\rho}^{z}\hat{G}_{0}(k)\hat{\rho}^{z}]^{-1},
\label{eq_ge}
\\&
\hat{G}_{\mathrm{O}}(k)=-\Delta\hat{G}_{0}(k)\hat{\rho}^{z}\hat{G}_{\mathrm{E}}(k).
\label{eq_go}
\end{align}
Here, $k=(\bm{k},i\omega_{n})$ with the fermionic Matsubara frequency, $\omega_{n}=(2n+1)\pi T$, and 
 we have used the facts that $\hat{\rho}^{z}\hat{G}_{0}$ and $\hat{M}$  
 are commutable, and $\hat{M}^{2}=\hat{1}$.  
 
From Eq.~(\ref{eq_g}), it is found that the angle-dependence of the AFM moment $\hat{M}(\phi')$ appears as the prefactor of $\hat{G}_{\mathrm{O}}(k)$. 
This $\phi'$-dependence  is reflected on the ME tensor through the atomic SOC.

\begin{figure}[t]
\begin{center}
\includegraphics[width=8cm]{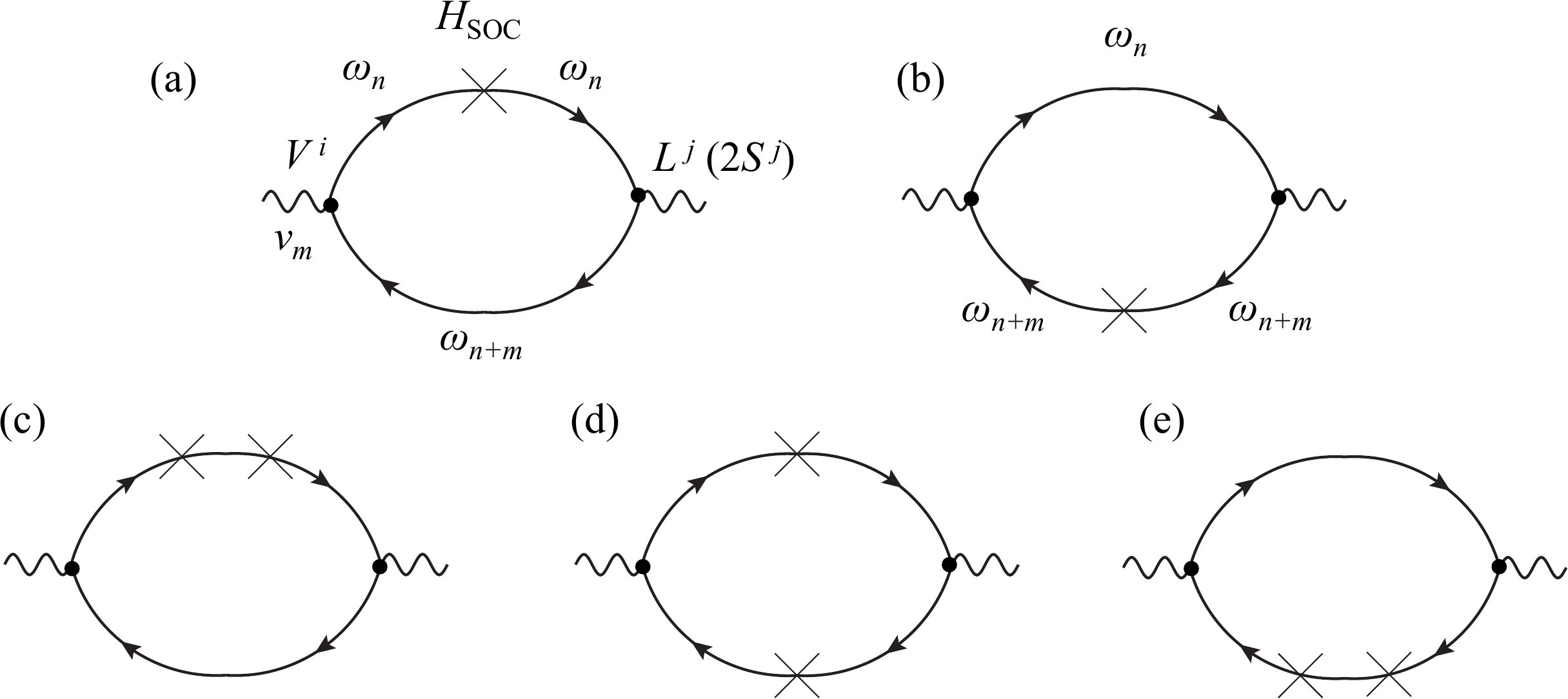}
\caption{The diagrammatic representation of the correlation function $Q^{L(S)}_{ij}(i\nu_m)$. 
The first and second order terms with respect to the atomic SOC are shown in  
(a) and (b), and (c)-(e), respectively. 
The arrows and crosses represent the non-perturbative Green's function $\hat{G}(k)$ and the SOC term $\mathcal{H}_{\mathrm{SOC}}$, respectively.
$Q^{L(S)}_{ij}(\omega+i\eta)$ is obtained by the analytic continuation, $i\nu_{m}\to\omega+i\eta$, in $Q^{L(S)}_{ij}(i\nu_m)$.
}
 \label{fig_diagram}
\end{center}
\end{figure}

Let us consider the first-order terms of the ME tensor with respect to the atomic SOC, $Q^{L(S)(1)}_{ij}(i\nu_m)$, which can be expressed 
 as products of $V^{i}$, $L^{j}(S^{j})$, $\mathcal{H}_{\mathrm{SOC}}$, and the three Green's functions $\hat{G}(k)$.
The corresponding diagrammatic representations are 
shown in Figs.~\ref{fig_diagram}(a) and (b).
 In what follows, the  symmetry arguments are useful to identify which perturbative terms remain finite. 
  For instance, the spin-diagonal  Green's function $\hat{G}_{\mathrm{E}}(k)$ is even-parity, while the spin-off-diagonal  
 $\hat{G}_{\mathrm{O}}(k)$ is odd-parity due to the additional 
 $\hat{\rho}^{z}$ in the latter.  
  Therefore, by considering the fact that $V^{i}$ is odd-parity, while $L^{i}$, $S^{i}$, and $\mathcal{H}_{\mathrm{SOC}}$ are even-parity,  
the  
perturbative terms containing  
odd numbers of $\hat{G}_{\mathrm{O}}(k)$ remain finite. 
 Moreover, $\mathcal{H}_{\mathrm{SOC}}$ in Eq.~(\ref{eq_hsoc}), $S^{i}$ in Eq.~(\ref{eq_s}), and $\hat{M}$
      in Eq.~(\ref{eq_g}) contain $\hat{\sigma}^{x}$ and $\hat{\sigma}^{y}$, but
       their products appeared in the perturbative terms 
       must be spin independent,  
       otherwise they vanish due to the trace over the spin indices. 
    As a result, $Q^{S(1)}_{ij}(i\nu_m)$ vanishes, while    
      $Q^{L(1)}_{ij}(i\nu_m)$ is given as 
  \begin{align}
&
\frac{Q^{L(1)}_{ij}(i\nu_{m})}{\lambda}= 
\left(I_{ij}^{x}+J_{ij}^{x}\right)\cos\phi'
+\left(I_{ij}^{y}+J_{ij}^{y}\right)\sin\phi',
\label{eq_pt_1} \\
&\quad
I_{ij}^{k}(i\nu_{m}) = -\int_{k}\mathrm{tr}\left[
\hat{v}^{i}(\bm{k})\hat{G}_{\mathrm{P}_1}(k_{+}) \hat{l}^{j} \hat{G}_{\mathrm{P}_2}(k) \hat{l}^{k}\hat{G}_{\mathrm{P}_3}(k)\right],
\label{eq_pt_2} \\
&\quad
J_{ij}^{k}(i\nu_{m}) = -\int_{k}\mathrm{tr}\left[
\hat{v}^{i}(\bm{k})\hat{G}_{\mathrm{P}_1}(k_{+}) \hat{l}^{k} \hat{G}_{\mathrm{P}_2}(k_{+}) \hat{l}^{j}\hat{G}_{\mathrm{P}_3}(k)\right],
\label{eq_pt_3}
\end{align}
 where we have introduced the abbreviation, $\int_{k}\equiv (T/N)\sum_{\bm{k}n}\sum_{\mathrm{P}_{1}\mathrm{P}_{2}\mathrm{P}_{3}}$ 
  where $\mathrm{P}_r$ takes either E or O, and $k_{+}=(\bm{k},i\omega_{n+m})$.
 As was mentioned, the only odd number of O in the summation ($\mathrm{P}_{1},\mathrm{P}_{2},\mathrm{P}_{3})$ gives finite contributions.
 The trace $\mathrm{tr}[\cdots]$ is taken over the orbital and sublattice indices. 
  Furthermore, the point-group argument concludes the following relations:  
 $-I_{xx}^{x}=I_{yy}^{x}=I_{xy}^{y}=I_{yx}^{y}=A(i\nu_{m})$, and the other components vanish.
 Similar relations also hold for $J^k_{ij}$.
 By these arguments, the first-order contribution is purely from the orbital part, and $\hat{\alpha}$ follows Eq.~(\ref{eq_me_exp}) with $r=1$.
 Similar arguments can be applied to  the second-order terms as shown in 
 Figs.~\ref{fig_diagram}(c)-(e).
It is found that the orbital contribution  vanishes, 
while $\hat{\alpha}^{S}$ contains both $\phi'$ and $3\phi'$ rotations, i.e., 
$r\ne 0$, $1$ in Eq.~(\ref{eq_me_exp}).

In summary, we have proposed a minimal model to exhibit the manipulating in-plane ME by extracting minimal ingredients from the realistic model
for Co$_4$Nb$_2$O$_9$.
The minimal conditions are (i) three $d$ orbitals in a trigonally distorted octahedron giving rise to the $xy$-type SOC, (ii) single honeycomb layer with 
 weak in-plane magnetic anisotropy, and (iii) weak SOC $\lambda$ as compared to AFM molecular field $\Delta$.
The $\bm{H}$-angle-dependence of the electric polarization in our minimal  
model is qualitatively consistent with experiments, and is understood by the perturbative argument with respect to $\lambda/\Delta$.
Our results can be applied to other AFMs, e.g., Co$_4$Ta$_2$O$_9$, showing the similar ME response.
These findings could be useful to explore efficiently further ME materials having similar manipulability of the electric polarization.

\begin{acknowledgments}
The authors would like to thank T. Arima  for fruitful discussions and directing our attention to the problem studied in the present work.
They also thank Y. Motome for many valuable discussions in the early stage of the present work. 
This work has been supported by JSPS KAKENHI Grant Number 15H05885 (J-Physics), 15K05176, and 16H06590.
\end{acknowledgments}


\begin{thebibliography}{99}
\bibitem{CEF_band}
In ref.~\cite{Yanagi}, the CEF splittings, $\varepsilon_{e_g}-\varepsilon_{a_{1g}}\sim 0.62$ ($0.84$)eV and $\varepsilon_{a_{1g}}-\varepsilon_{e_{g}'}\sim 0.09$ ($0.15$)eV 
 at a Co(1) [Co(2)] site have been obtained from the band calculation.
These values are somewhat different from those in ref.~\cite{Solovyev}. 
This is mainly because the different structural parameters were used, namely, ref.~\cite{Castellanos} was 
used in ref.~\cite{Solovyev}, and \cite{Bertaut} was used in ref.~\cite{Yanagi}.
\end{thebibliography}
\end{document}